\documentclass[
    ,final            
  ]
  {aipproc}

\layoutstyle{8x11double}

\begin{document}

\title{Review of overall parameters of giant radio pulses
from the Crab pulsar and B1937+21}


\author{Bilous A.V.}{
  address={Astro Space Center of Lebedev Physical Institute, Moscow, Russia}
  ,altaddress={Moscow Institute of Physics and Technology}
}

\author{Kondratiev V.I.}{
  address={Astro Space Center of Lebedev Physical Institute, Moscow, Russia}
 ,altaddress={West Virginia University, Morgantown, USA}
}

\author{Popov M.V.}{
 address={Astro Space Center of Lebedev Physical Institute, Moscow, Russia}
}

\author{Soglasnov V.A.}{
 address={Astro Space Center of Lebedev Physical Institute, Moscow, Russia}
}

\begin{abstract}
 We present a review of observed parameters of
 giant radio pulses, based on the observations conducted by our group during recent
 years.
 The observations cover a broad frequency range of about 3  octaves,
 concentrating between 600 and 4850 MHz. Giant pulses of both the
 Crab pulsar and the millisecond pulsar B1937+21 were studied with the
 70-m Tidbinbilla, the 100-m GBT, 64-m Kalyazin and Westerbork radio telescopes.  We discuss pulse energy distribution, dependence of peak flux density
 from the pulse width, peculiarities of radio spectra, and polarization
 properties of giant radio pulses.

\end{abstract}

\maketitle

\section{Introduction}
Giant pulses (GPs) are abnormally    
short  (down to nanoseconds), bright (up to MJy), polarized (up to 100\%)
bursts of pulsar radio emission. They occur in the very narrow windows 
and are known only for a handful of pulsars.

Here we present the detailed study of two brightest sources of giant 
pulses -- the Crab pulsar and millisecond pulsar B1937+21.

Both the Crab pulsar and the PSR B1937+21 have many similar characteristics 
that can be a clue for understanding the giant pulse phenomenon. 
They both have an interpulse in their regular  profiles and  highest 
values of magnetic field at the light cylinder (about $10^6$~G). Popov et al.\cite{MV2006}  
have recently proposed that the Crab regular profile consists of only 
GPs whereas the precursor observed at the frequencies below 1 GHz represents 
the regular profile. Hence, for both the Crab pulsar and the pulsar 
B1937+21 GPs occur at the very trailing edge of the regular emission profile.

Polarization properties are also similar. In general, GPs from the Crab 
pulsar are wider (up to $100~\mu$s), mainly with some degree of linear polarization. 
However, Hankins et al. \cite{Hankins}  showed that these pulses consist 
of very narrow (<2 ns) bursts with 100\% circular polarization, 
either left or right. The number of these fundamental bursts in any 
individual broad GP was estimated by  Popov et al.\cite{MV2006} 
to be about 100. Giant pulses from the B1937+21 reveal the broad 
spectrum of polarization properties: they can be almost 100\% circular 
polarized, and they may have high linear polarization as well. This means 
that probably they also consist of narrower pulses. 

Many similar properties are undoubtedly manifestations of the same physical 
process responsible for GP emission. However, there are still some differences, 
that represent either the individual character of these two pulsars or the influence of the ISM, or both.

\section{Energy distribution}

Unlike regular emission, which has normal  
or log-normal energy distribution, giant 
pulses follow  power-law energy distribution:
\begin{equation}
N (\mbox{GPs with }E > E_0) \sim  {E_0}^\alpha
\end{equation}

Both for the Crab pulsar and B1937+21 there is a correlation between GP energy and width:
the brightest pulses have the shortest duration. As a consequence, giant pulses with
different duration have distributions with different values of $\alpha$ and the distribution
in GP width depends on the range of energy used.

One can raise the question: is there any cutoff
of power-law distribution at low and at high energies?

The recent 160 hours monitoring campaign\footnote{Simultaneous observations at 600, 1650
and 4850~MHz at Kalyazin radio telescope with time resolution 500 $\mu$s. Both circular
polarizations were recorded at 1650 and 4850 MHz, only RCP -- at 600 MHz.} of the Crab pulsar carried out by 
Popov et al.\cite{MV2007} revealed that the 
energy distributions of GP \begin{bf}do follow a power-law\end{bf} at both 600 and 4850 MHz 
\begin{bf}up to the highest energy detected\end{bf}, namely 5 MJy$\cdot\mu$s for 600~MHz and 0.07 MJy$\cdot\mu$s
for 4850~MHz.

The investigation of energy distribution of Crab GPs with respect to their different 
widths carried out by Popov \& Stappers \cite{Stappers}  at 1200 MHz\footnote{Westerbork 
Synthesis RT, total intensity recorded for 3.5 hr with time reolution of 4.1 $\mu$s.} can shed light on
the low-energy part of the distribution. It was disclosed 
that the distributions manifest notable differences for the different width groups.
For main pulse, the wider are GPs, the steeper is their distribution ($\alpha$ varying
from $-1.7$ to $-3.2$). There are also 
\begin{bf}breaks in the power-law fits indicating flattening at low energies\end{bf} ($\alpha$ switches to
 $-1.0$ - $-1.9$). At the same time, the interpulse GPs can be fitted with one 
 power-law function with $\alpha = -1.65$.
\begin{figure}
\includegraphics[height=.3\textheight,angle=-90]{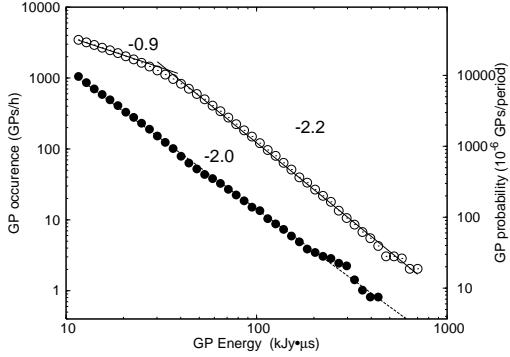}
\caption{Cumulative energy distribution for the Crab pulsar GPs at 600 MHz.
Filled circles mark interpulse, open -- main pulse.}
\label{Crab600}
\end{figure}
A similar situation was found at 600 MHz\footnote{Kalyazin RT, 295-min observation with RCP recorded with time resolution of 125 ns.} 
(see Figure \ref{Crab600}), 
although in this case GPs from the interpulse are fitted by  power-law with $\alpha$ varying from $-1.6$ to $-3.1$ 
going from the shortest to the longest GP.  

The break-point pulse energy found (4~kJy$\cdot \mu\mbox{s}$ at 1200 MHz and 40~kJy$\cdot \mu\mbox{s}$
at 600 MHz) suits a power-law frequency dependence 
presented by  Popov \& Stappers \cite{Stappers}:
\begin{equation}
 E_{break}  ( f ) = 7 f^{-3.4} \qquad \mbox{kJy$\cdot \mu\mbox{s}$,}
 \end{equation}
 where $f$ is in GHz.

Giant pulses from B1937+21 are in general fainter than those from the Crab pulsar,
so it is more difficult to collect the sufficient number of events.

As a result of observation of  B1937+21 at the 70-m Tidbinbilla radio telescope at 
1650 MHz\footnote{RCP recorded for 39 min with time resolution of 31.25 ns.}
we found that GP with energies from 100 to 1000 Jy$\cdot \mu$s follow a power law
with $\alpha=-1.4$ for both main pulse and interpulse.

\begin{figure}
\includegraphics[height=.3\textheight,angle=-90]{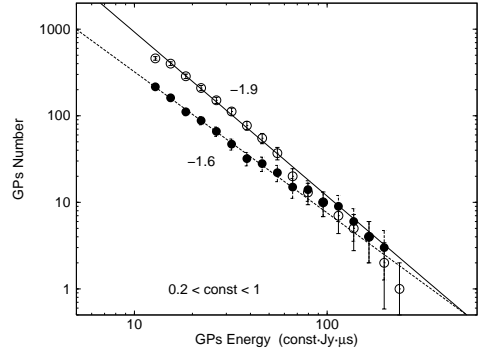}
\caption{Cumulative energy distribution for GPs from B1937+21 at 2100 MHz.
Main pulse is marked by open circles, interpulse -- by filled circles.}
\label{B37_2100}
\end{figure}

Another attempt to found low-energy break was made at 2100 MHz using the
100-m Green Bank telescope\footnote{Both circular polarizations were recorded for 7.5 hr 
with time resolution of 7.8 ns.}. 
The result is shown in Figure \ref{B37_2100}.
Owing to much better GBT sensitivity, 
we were able to detect more low-energy GPs at 2100 MHz, but we still 
did not find the low-energy cutoff. 

\section{Polarization properties}

As the result of observations of B1937+21 at 2100 MHz$^5$ (see \cite{Vlad} for details)
we found that
the majority of GPs (>55\%) have circularly polarized peaks with 
fractional polarization >0.8 (either left or right). The fractional 
linear polarization of GPs is also very high, namely 48\% have fractional 
linear polarization of 0.4-0.5.

The position of the peak in circular and linear polarization profiles 
within the same GP could be different. Hence, the same GP can reveal 
both strong circular and strong linear polarization. The fraction of 
GPs with both large circular (>0.8) and linear (>0.4) fractional 
polarizations is almost 38\%. Thus, GPs from B1937+21 are very strongly 
polarized, both circularly and linearly.

\begin{figure}
\includegraphics[height=.3\textheight,angle=-90]{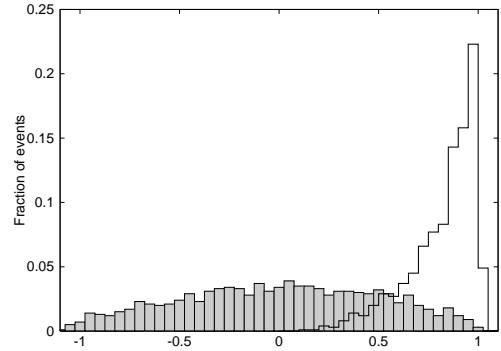}
\caption{Distribution of polarization degree of  GPs from Crab pulsar at 2244 MHz.
Empty bars represent the degree of linear polariztion, filled -- circular 
(negative for LCP and positive for RCP).}
\label{Crab_pol}
\end{figure}

Polarization properties of the Crab GPs\footnote{Kalyazin RT, 3 hours of 
recording both RCP and LCP at 2244 MHz with time resolution of 31.25 ns.}  
 are displayed in  Figure \ref{Crab_pol}.
In general, the properties 
are identical: high degree of linear and circular polarization with 
random fluctuations in different spikes within the same giant pulse.

\section{Pecularities of radio spectra}

\begin{figure}
\includegraphics[height=.4\textheight]{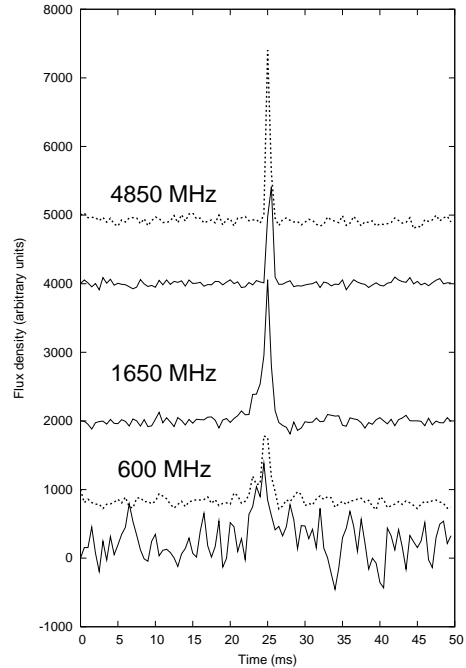}
\caption{An example of GPs found simultaneosly on 600, 1650 and 4850 MHz.
Solid line stands for RCP, dotted -- for  LCP.}
\label{GP}
\end{figure}

 Eilek \& Hankins 
\cite{Eilek}  revealed that interpulse GPs at frequencies more than 3 GHz 
differ significantly from GPs at main pulse both in profile and 
spectrum properties. The spectra of high-frequency interpulse GPs 
show striking regularly spaced structure, while those from main 
pulse do not. Hence, the hypothesis was proposed that interpulse 
GPs at frequencies > 3 GHz originate in another region inside the 
pulsar magnetosphere. During our tri-frequency monitoring of the Crab pulsar$^1$ 
we have found 8 GPs on 
interpulse longitudes occurred simultaneously at 3 frequencies: 600, 
1650, and 4850 MHz (see Figure \ref{GP}). The probability of an occasional coincidence of 2 
independent events in the same pulse is 0.0003. Thus, the suggestion 
about other localization of high-frequency interpulse GP emitters 
is very unlikely.

In addition, we have found the large number of GPs that occurred simultaneously at 600 
and 4850 MHz, but not at 1650 MHz, so 
GP spectra must have large-scale irregularities with the scale 
$\delta \nu / \nu$ of about 0.5.  
 
It was revealed (see \cite{MV2007} for details), that GPs at high frequency have flatter indices
(even positive) than those at low frequency. In general, radio spectra 
in this range can not be represented by an unique power-law function, 
or even power-law function with break. 

\section{Discussion}
It seems rather unlikely that any plasma mechanism is responsible for GP emission. Extermely short GP duration implies very small GP emission region, $10^2 - 10^3$ cm. For strongest GPs, volume density of the emission exceeds by many orders density of plasma energy, it is comparable with the density  of the energy of the magnetic field. Huge peak flux densities for the strongest GPs from the Crab pulsar detected at the Earth means an ultimate field strength of EM wave inside pulsar's magnetosphere.
Interaction of such strong wave with plasma is very specific: the wave accelerates charged particles up to relativistic speed over one wave cycle, and the particles emit secondary waves at different frequency in different directions than the incident wave. The process may explain unusual HFC1 and HFC2 components in the profile of the Crab pulsar.

\bibliographystyle{aipproc}   

\bibliography{articles}

\end{document}